\documentclass[twocolumn,showpacs,superscriptaddress,amssymb,10pt,pra,floatfix,aps,longbibliography]{revtex4-2}

\usepackage{graphicx}
\usepackage{amsmath}
\usepackage{dcolumn}
\usepackage{bm}
\usepackage{epsfig}
\usepackage{color}
\usepackage{longtable}
\usepackage{leftidx,amsmath}
\usepackage{natbib} 
\usepackage[utf8]{inputenc}
\usepackage{braket}
\usepackage{tikz}
\usepackage{makecell}
\usepackage[flushleft]{threeparttable}
\usepackage[utf8]{inputenc}
\usepackage{multirow}
\usepackage{isotope}
\usepackage{array}
\usepackage{siunitx}

\usepackage{orcidlink}
\usepackage{url}

\usepackage{dcolumn}
\newcolumntype{d}{D{.}{.}{-1}}
\newcolumntype{f}[1]{D{.}{.}{#1}}

\newcommand{\eg}{{e.g., }}

\def\emc2{\ensuremath{~m_\text{e}c^2}}

\begin{document}
	\preprint{}
	%
%
%
%
	\title{All-Order Wichmann and Kroll Contribution in Heavy Electronic and Exotic Atoms}
%
%
%
%

\author{Jonas Sommerfeldt\orcidlink{0000-0002-3471-7494}}
	\email{jonas.sommerfeldt@lkb.upmc.fr}
	\affiliation{Laboratoire Kastler Brossel, Sorbonne Universit\'e, CNRS, ENS-PSL Research University, Coll\`ege de France, Case\ 74;\ 4, place Jussieu, F-75005 Paris, France}
    
\author{Paul Indelicato\orcidlink{0000-0003-4668-8958}}\email{paul.indelicato@lkb.upmc.fr}
\affiliation{Laboratoire Kastler Brossel, Sorbonne Universit\'e, CNRS, ENS-PSL Research University, Coll\`ege de France, Case\ 74;\ 4, place Jussieu, F-75005 Paris, France}

	\date{\today \\[0.3cm]}

%
%
\begin{abstract}
We present a theoretical study of the Wichmann and Kroll correction to the one-loop vacuum polarization (VP) to all-orders in $\alpha Z$. We consider electronic, muonic, and antiprotonic atoms for a wide range of nuclear charge numbers and explicitly investigate the influence of finite nuclear size effects and different nuclear models. Moreover, we place special emphasis on circular Rydberg states in the exotic atoms as they have recently attracted interest as a tool to perform isolated tests of strong-field QED. We find that the higher-order vacuum polarization is strongly enhanced in exotic atoms and remains large enough in highly excited Rydberg states that an accurate treatment is crucial for the analysis of upcoming spectroscopy experiments like PAX. Moreover, our calculations show that the VP contribution to the Lamb shift in these exotic Rydberg states has almost no dependence on the structure of the nucleus.
\end{abstract}
\maketitle


\section{\label{sec:introduction} Introduction}
Quantum electrodynamics (QED) has been tested to extraordinary precision and is indispensable for tests of the Standard Model at the precision frontier. One of the most important QED effects, the Lamb shift of atomic energy levels, has been measured in hydrogen with a relative accuracy around $10^{-9}$ \cite{pmab2011,mppa2013} and is in perfect agreement with the most precise theoretical predictions~\cite{ypp2019}. However, tests of QED in atoms with a high nuclear charge number $Z$ are orders of magnitude less accurate despite the fact that the strong fields in these systems allow to probe a qualitatively different sector of the Standard Model: In the high-$Z$ regime, the commonly used $\alpha Z$ expansion, where $\alpha \approx 1/137$ is the fine-structure constant, does not converge and a non-perturbative treatment of the interaction of the quantum vacuum with the Coulomb field of the nucleus is necessary~\cite{mps1998, ind2019}. Apart from experimental and theoretical difficulties, one key factor that obstructs precision tests of QED in high-$Z$ systems are uncertainties that arise from imprecise knowledge of nuclear properties, which limits the ability to accurately predict transition energies in heavy atomic systems. However, it has been shown that in exotic atoms, where an electron in the atomic shell has been replaced by a heavier exotic particle, the transitions between Rydberg states are almost unaffected by nuclear effects, whereas QED contributions remain relatively large, allowing for an almost isolated test of strong-field QED~\cite{pba2021}. This approach will be utilized by several experiments, such as the PAX experiment, which will use antiprotons from the ELENA ring at CERN~\cite{brr2025} and HEATES which performs spectroscopy with muonic atoms at the J-PARC facility~\cite{oabc2020, oabc2023}. However, to reach the full potential of these experiments as QED precision tests, their accuracy needs to be matched by theory. 

In contrast to electronic atoms, the Lamb shift in exotic atoms is mostly dominated by the vacuum polarization (VP) contribution, whereas the effects of the self-energy (SE) are almost negligible. While perturbative calculations (in $\alpha Z$) of the leading order VP are relatively easy to perform~\cite{ueh1935, wak1956}, all order calculations need to be carried out in the framework of bound-state QED and pose a greater challenge due to the more complicated structure of the Dirac-Coulomb propagator. The first calculations of the all-order VP was performed by Rinker and Wilets~\cite{raw1973} and Gyulassy~\cite{gyu1974, gyu1974a, gyu1975}, which were later improved with a different approach based on the partial-wave expansion of the electron propagator~\cite{sam1988, plss1993}. Since then, many studies have performed similar calculations in electronic atoms with higher accuracy and more realistic nuclear models~\cite{fmn1991, plss1993, asy1997, yer2011, sas2023, ibgv2024}. In contrast, exotic atoms have received less attention. While some studies considered heavy muonic atoms~\cite{blo1972, raw1975, ras1977, bar1982, ssm1989} and recently, Patkó$\Check{\text{s}}$ and Pachucki have performed perturbative (in $\alpha Z$) VP calculations in antiprotonic atoms~\cite{pap2025}, to our knowledge, exact relativistic calculations of the all-order vacuum polarization of antiprotonic atoms are not available~\cite{kik2007, pba2021, brr2025}. 

In this contribution, therefore, we present a theoretical study of the all-order Wichmann and Kroll correction in heavy electronic and exotic atoms. A special emphasis is placed on highly excited circular  Rydberg states and on the influence and uncertainty coming from the structure of the nucleus. We start our analysis by discussing the formalism and numerical methods to calculate the VP potential to all orders in $\alpha Z$ in the framework of bound-state QED in Section~\ref{Sec:VPPot}. After that, we will use these results to calculate the VP contribution to the Lamb shift in electronic, muonic, and antiprotonic atoms in Section~\ref{Sec:VPLamb}. Finally, in Section~\ref{Sec:Conc}, we will summarize our results. Relativistic units ($\hbar = c = m_e = 1$, $\alpha = e^2$) are used in this paper if not stated otherwise.

\section{One-Loop Vacuum Polarization Potential} \label{Sec:VPPot}
\begin{figure*}
    \centering
      \includegraphics[width=0.85\textwidth]{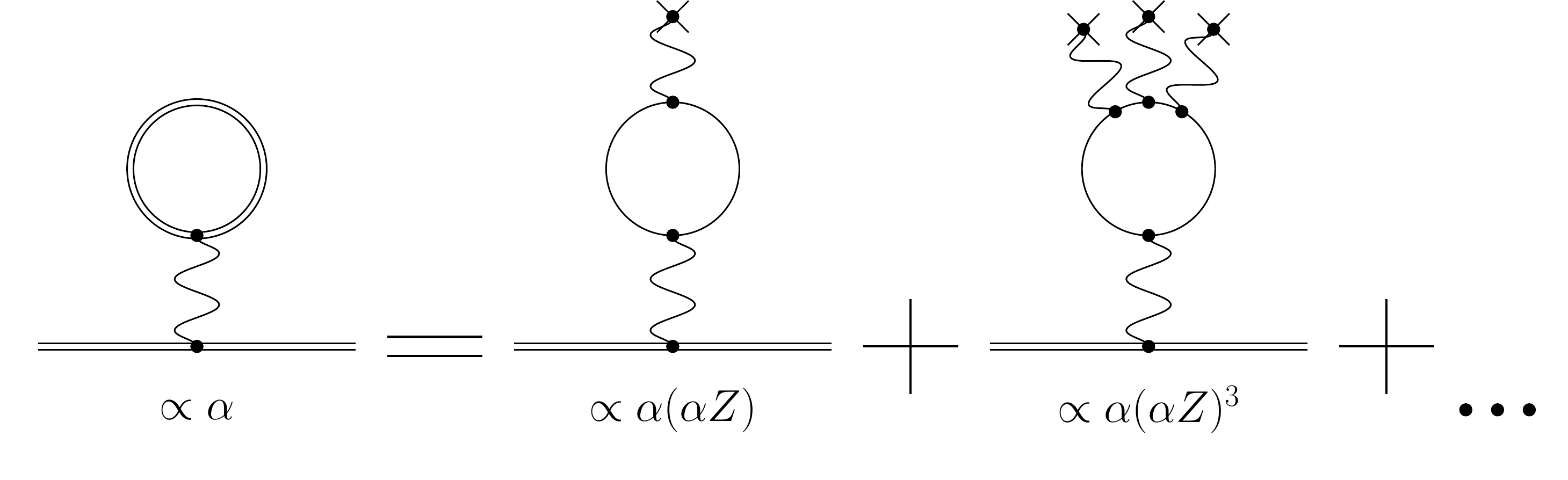}
       \caption{Leading-order in $\alpha$ and all-order in $\alpha Z$ Feynman diagram for the vacuum polarization contribution to the Lamb shift (left side of the equation) and its expansion into powers of $\alpha Z$ (right side of the equation). \label{Fig:VPFeynman}}
\end{figure*}

The all-order VP contribution to the Lamb shift can be represented by the Feynman diagram on the left side of Fig.~\ref{Fig:VPFeynman} with its expansion into powers of $\alpha Z$ is shown on the right side of the figure. It can be shown that the VP contribution to the Lamb shift can be described by a scalar potential
\begin{equation}
     V_{VP}(r) = V_{VP}^{(1)}(r)+ V_{VP}^{(3+)}(r)~,
\end{equation}
where $V_{VP}^{(1)}(r)$ corresponds to the lowest order in $\alpha Z$, the so-called Uehling term, and $V_{VP}^{(3+)}(r)$ is the all-order Wichmann and Kroll contribution that contains all terms with 3 or more interactions with the Coulomb potential. It is convenient to evaluate these two terms separately as only the Uehling term is divergent while the remaining higher-order part is finite.

\subsection{Uehling Contribution}
The effect of the first-order VP correction can be expressed by the potential
\begin{equation} \label{Eq:VVP1}
\begin{aligned}
    V_{VP}^{(1)}(r) = -e\frac{2\alpha}{3} \frac{1}{r} &\int_0^\infty dr'~r' \rho(r')\\
    &\times [\chi_2(2\vert r - r' \vert)-\chi_2(2(r + r'))] ~,
\end{aligned}
\end{equation}
where 
\begin{equation}
    \chi_n(z)=\int_1^\infty dt~e^{-tz}t^{-n}\left(1+\frac{1}{2t^2}\right)\sqrt{1-\frac{1}{t^2}}~,
\end{equation}
and $\rho(r)$ is the nuclear charge distribution which assumed to spherially symmetric (see, e;g., \cite{kla1977}). In this work, we consider four nuclear models: The point-, shell-, sphere- and Fermi-model. The nuclear potential $V(r)$ can be obtained from the respective charge distributions
\begin{align} \label{Eq:NucM}
    \begin{split}
        \rho_\text{point}(r) &= \rho_0~\!\delta(r)/r^2~, \\
        \rho_\text{shell}(r) &= \rho_0~\!\delta(R - r)~, \\
        \rho_\text{sphere}(r) &= \rho_0~\!\Theta(R_\text{sphere} - r)~, \\
        \rho_\text{Fermi}(r) &= \frac{\rho_0}{1+\exp[(r-c)/a]}~,
    \end{split}
\end{align}
where the parameters $a$ and $c$ of the Fermi distribution are given by $a = t/(4\ln 3)~\text{fm}$ and $c^2 = \frac{5}{3}R^2 - \frac{7}{3}a^2\pi^2$ with $R$ being the root-mean-square radius of the nucleus. In this work we use $t=\qty[mode =text]{ 2.3}{fm}$.  The radius of the sphere distribution is given by $R_\text{sphere} = \sqrt{5/3}R$.  Moreover, $\Theta(r)$ is the Heaviside step function and $\rho_0$ is a normalization constant that is chosen so that $4\pi\int dr~r^2 \rho(r)=eZ$, where $Z$ is the proton number of the nucleus.
\\

\subsection{All-Order Wichmann and Kroll Contribution}
We will now turn to the evaluation of the remaining higher-order VP contribution. Following Refs.~\cite{sam1988,ims2014}, we can conveniently express the all-order VP potential as
\begin{equation} \label{Eq:AOVP}
    V_{VP}(\boldsymbol{r}) = -e\int~d^3\boldsymbol{r'}~\frac{\rho_{VP}(\boldsymbol{r'})}{\vert \boldsymbol{r} - \boldsymbol{r'} \vert}~,
\end{equation}
where 
\begin{equation}
    \rho^{\vphantom{(1)}}_{VP}(\boldsymbol{r}) = \frac{e}{2\pi i}\int dz~\text{Tr} [G(\boldsymbol{r}, \boldsymbol{r}, z)]
\end{equation}
is the all-order VP charge density and $G(\boldsymbol{r}_2, \boldsymbol{r}_1, z)$ is the Dirac-Coulomb Green's functions, which includes all interactions of the virtual electron-positron pair with the Coulomb field of the nucleus. Since the Uehling contribution is divergent, directly evaluating Eq.~\eqref{Eq:AOVP} will also yield a divergent result. However, it can be easily shown that the difference between the all-order and lowest-order diagrams gives a finite result for the desired part of the VP with 3 or more interactions. Subtracting the lowest-order term and choosing an integration contour along the imaginary axis for the integral over the energy z yields
\begin{widetext}
\begin{equation} \label{Eq:VP3+}
\begin{aligned}
    \rho_{VP}^{(3+)}(r) =& \rho^{\vphantom{(1)}}_{VP}(r) - \rho_{VP}^{(1)}(r) \\
    =& \frac{e}{2\pi^2} \int_0^\infty du~ \left\{ \sum_{\kappa = \pm 1}^{\pm \infty} \vert \kappa \vert~\text{Re}\left[ \sum_{i=1}^2 G_\kappa^{ii}(r, r, iu) + \int_0^\infty dr'~r'^2V(r') \sum_{n,m=1}^2 [F^{nm}_\kappa(r,r',iu)]^2\right]\right\} \\
    &+\frac{e}{2\pi} \sum_{-m<E_{n\kappa}<0} \vert \kappa \vert \left[g^2_{n\kappa}(r) + f^2_{n\kappa}(r) \right]~,
\end{aligned}
\end{equation}
\end{widetext}
where $F_\kappa^{}(r_1, r_2, z)$ and $G_\kappa^{}(r_1, r_2, z)$ are the Green's function for the free radial Dirac equation and for the radial Dirac equation that includes the nuclear potential $V(r)$, respectively. Moreover, $g_{n\kappa}$ and $f_{n\kappa}$ are the large and small components of the bound solutions of the Dirac equation with principal quantum number $n$, Dirac quantum number $\kappa$ and eigenenergy $E_{n\kappa}$. It can be shown that the spurious part of the third-order term in Eq.~\eqref{Eq:VP3+} vanishes if the summation over $\kappa$ is performed over a finite number of terms and no further subtraction is necessary in this case~\cite{sam1988}. The most difficult part in the evaluation of Eq.~\eqref{Eq:VP3+} is calculating the Green's function $G_\kappa^{}(r_1, r_2, z)$ in a numerically stable way for arbitrary potentials and arguments.

The radial Dirac-Coulomb Green's function can be expressed as
\begin{equation}
\begin{aligned}
G_\kappa(&r_2,r_1,z) = \frac{1}{w_\kappa(z)} \\
&\times\Bigg[\Theta(r_2-r_1) \left(\begin{array}{c}
F_{\kappa,\infty}^1 (r_2,z)\\
F_{\kappa,\infty}^2 (r_2,z)\\
\end{array}\right) \left(\begin{array}{c}
F_{\kappa,0}^1 (r_1,z)\\
F_{\kappa,0}^2 (r_1,z)\\
\end{array}\right)^T\\
&+\Theta(r_1-r_2) \left(\begin{array}{c}
F_{\kappa,0}^1 (r_2,z)\\
F_{\kappa,0}^2 (r_2,z)\\
\end{array}\right) \left(\begin{array}{c}
F_{\kappa,\infty}^1 (r_1,z)\\
F_{\kappa,\infty}^2 (r_1,z)\\
\end{array}\right)^T\Bigg]~,
\end{aligned}
\end{equation}
where
\begin{equation} \label{Wronskian}
w_\kappa(z) = r^2 [F^2_{\kappa,0}(r,z) F^1_{\kappa,\infty}(r,z)-F^1_{\kappa,0}(r,z) F^2_{\kappa,\infty}(r,z)]~,
\end{equation}
is the Wronskian, and $F^{1,2}_{\kappa,0}(r)$ and $F^{1,2}_{\kappa,\infty}(r)$ are the solutions of the homogeneous Dirac equation
\begin{equation}
\left(\begin{array}{cc}
1 + V(r) - z & -\frac{1}{r} \frac{\text{d}}{\text{d}r} r + \frac{\kappa}{r}\\
\frac{1}{r} \frac{\text{d}}{\text{d}r} r + \frac{\kappa}{r} & -1 + V(r) - z
\end{array}\right)\left(\begin{array}{c}
F_{\kappa}^1 (r,z)\\
F_{\kappa}^2 (r,z)\\
\end{array}\right)  = 0 ~,
\end{equation} 
that are regular at the origin and at infinity, respectively, see Ref.~\cite{mps1998} for further details. 

If the nuclear potential $V(r)$ is assumed to be a pure Coulomb potential of a point charge, closed analytical solutions can be found in terms of Whittaker functions~\cite{mps1998, hyl1984}. For arbitrary potentials, we employ the method of Salvat and co-workers~\cite{sam1991} and solve the Dirac equation on a grid between the origin and a point $R_0$, after which the potential is assumed to be a Coulomb potential. For the solutions regular at the origin $F_0$, we integrate outward from $r = 0$ to $r = R_0$. For $r > R_0$, $F_0$ is given by
\begin{equation}
    F_{\kappa, 0}(r,z) = aF_{\kappa, 0}^C + bF_{\kappa, \infty}^C~,~r\geq R_0~,
\end{equation}
where $F_{\kappa, 0}^C$ and $F_{\kappa, \infty}^C$ are the analytical Dirac-Coulomb solutions and the coefficients $a$ and $b$ are determined by the requirement that both components are continuous at $r = R_0$. The solution regular at infinity $F_\infty$ is equal to the analytical Dirac-Coulomb solution for $r \geq R_0$:
\begin{equation}
    F_{\kappa, \infty}(r,z) = F_{\kappa, \infty}^C~,~r\geq R_0~.
\end{equation}
To obtain the solution in the inner region, we solve the Dirac equation inward from $r = R_0$ to the origin, see Appendix~\ref{Ap:Dirac}.

Having discussed the evaluation of the Green's function $G_\kappa(r_1, r_2, z)$, we will now briefly cover the subtraction of the divergent lowest-order term in Eq.~\eqref{Eq:VP3+}. For the shell- and sphere nuclear models, the potential can be written in a form $V(r) = V_0 + V_2r^2$ in the inner region and as a Coulomb potential in the outer one. Therefore, the integral
\begin{equation} \label{Eq:OITerm}
    \int_0^\infty dr'~r'^2V(r') \sum_{n,m=1}^2 [F^{nm}_\kappa(r,r',iu)]^2~,
\end{equation}
can be solved analytically by adapting the method used by Soff and Mohr~\cite{sam1988} for the shell nucleus to the more general case, see Appendix~\ref{Ap:I1Ana}. For the Fermi model, we numerically integrate Eq.~\eqref{Eq:OITerm} up to $R_0$ and use the analytical solution for the integral from $R_0$ to infinity. Lastly, for the case of a point-like nucleus, we use a different approach that is faster and numerically more stable by analytically differentiating the Dirac-Coulomb Green's function instead of calculating the integral~\eqref{Eq:OITerm} to obtain the term with one interaction, see Ref.~\cite{yas1999}.

\begin{figure*}
    \centering
      \includegraphics[width=0.92\textwidth]{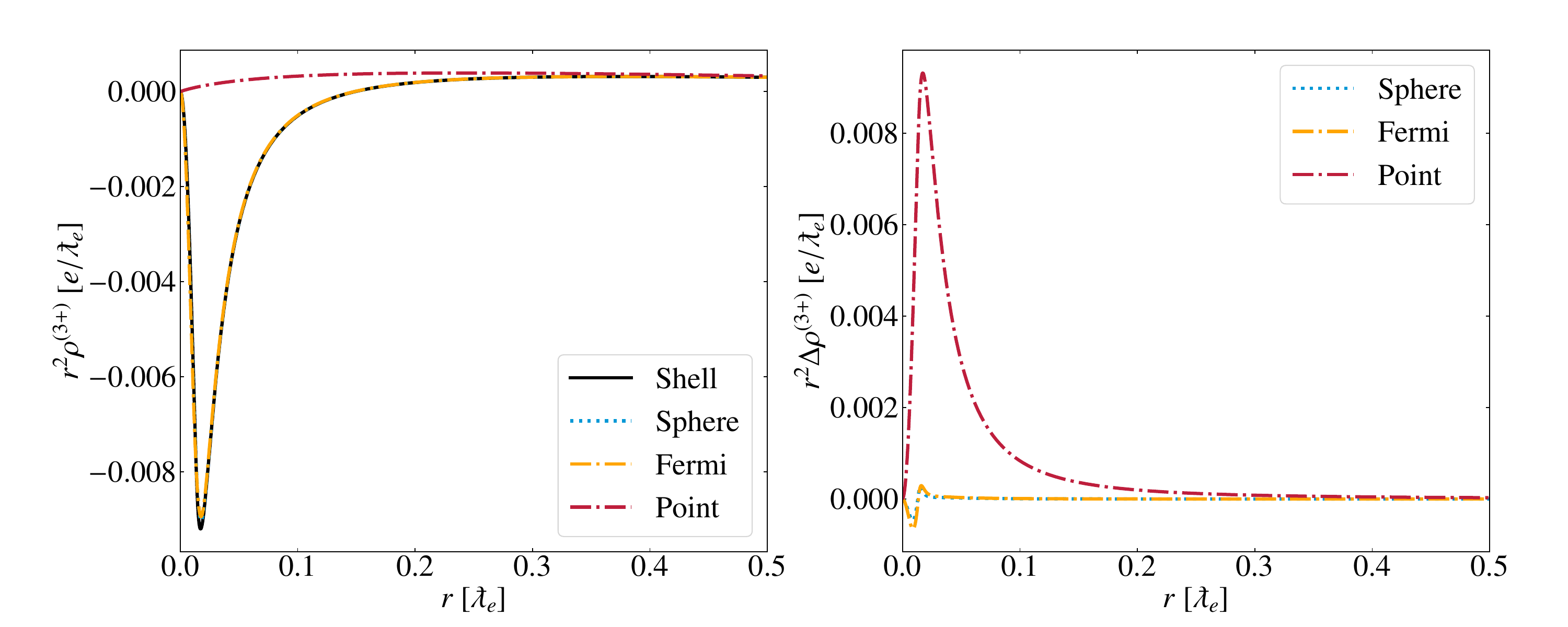}
       \caption{Higher-order vacuum polarization charge distribution~\eqref{Eq:VP3+} (left panel) for \isotope[208][82]{Pb} and shell (black solid line), sphere (blue dotted line), Fermi (orange dash-dotted line) as well as point (red dashed line) nuclear models. The differences of the nuclear models to the shell model are shown on the right panel. The summation over $\kappa$ was carried out until $\kappa = \pm 12$~. \label{Fig:VPCD}}
\end{figure*}

We are now finally ready to evaluate the VP charge distribution~\eqref{Eq:VP3+} for arbitrary nuclear potentials. In Fig.~\ref{Fig:VPCD} we show the model dependence for a bare lead nucleus. 
As seen from the figure, the VP charge is strongly localized near the nucleus. Moreover, the charge distribution of the point-like and finite nuclear models only differ significantly close to and inside the nucleus, while the shell-, sphere- and Fermi-models are very similar for all radial coordinates. In practical calculations of the VP potential, we follow Ref.~\cite{sam1988} and evaluate $\rho^{(3+)}_{VP}(r)$ between $r = 0$ and $r = 500R$. Using the fact that the total induced VP charge $Q = 4 \pi \int dr~r^2 \rho^{(3+)}_{VP}(r)$ must be zero, we can extrapolate the charge distribution to large distances to increase the numerical accuracy by assuming that it decreases as an inverse power of $r$. 

The relative theoretical uncertainty that results from the extrapolation of the charge distribution as well as the numerical methods used to obtain the Dirac-Coulomb Green's function and calculate the integrals in Eqs.~\eqref{Eq:AOVP} and \eqref{Eq:VP3+} is below $10^{-5}$. The largest source of uncertainty is the partial wave expansion in Eq.~\eqref{Eq:VP3+}. Our calculations have shown that summing up to $\kappa = \pm 12$ results in a relative uncertainty on the order of $10^{-5}$ for electronic atoms. The convergence is more rapid for exotic particles due to the smaller Bohr radius where higher-order multipoles contribtute less.

\section{Vacuum Polarization Contribution to the Lamb Shift} \label{Sec:VPLamb}
\subsection{Electronic Atoms}
We will now briefly discuss the all-order Wichmann and Kroll shift in electronic atoms before considering their exotic counterparts. The one-loop VP energy shift can be obtained from the integral
\begin{equation} \label{Eq:EVP}
   E_{VP} = \int d^3 \boldsymbol{r}~\psi^\dagger(\boldsymbol{r}) V_{VP}(r) \psi(\boldsymbol{r})~,
\end{equation}
where $\psi(\boldsymbol{r})$ is the electron wave function for the state under consideration which is given by the solution of the Dirac equation with the respective nuclear potential included. The VP potential $V_{VP}(r)$ in Eq.~\eqref{Eq:EVP} is calculated from Eqs.~\eqref{Eq:VVP1} and using
\begin{equation}
\begin{aligned}
    V_{VP}^{(3+)}(r) = -4\pi e&\left[ \frac{1}{r} \int_0^{r'} dr'~r'^2\rho^{(3+)}(r') \right. \\
    &\left.+ \int_{r'}^\infty dr'~r'\rho^{(3+)}(r')\right]~,
\end{aligned}
\end{equation}
for the contributions of Uehling and higher-order Wichmann and Kroll, respectively. We display the results for the one-loop VP with 3 or more interactions in Table~\ref{Tab:MDEl} for hydrogen-like \isotope[208][82]{Pb} and \isotope[238][92]{U} and the four nuclear models from Eq.~\eqref{Eq:NucM}.
\begin{table}
\begin{ruledtabular}
\begin{tabular}{ccccc}
\\[-8pt]
State & Point & Shell & Sphere & Fermi\\
\\[-8pt]
\colrule
\\[-6pt]
\multicolumn{5}{c}{\isotope[208][82]{Pb}}\\
\\[-6pt]
$1s_{1/2}$ & 2.4153\hphantom{$^{\ref{FN:TabRef}}$} & 2.2892 & 2.2898 & 2.2901\hphantom{$^{\ref{FN:TabRef}}$}\\
& 2.4155\footnote{\label{FN:TabRef}Ref.~\cite{yas2015}} & & & 2.2906$^{\ref{FN:TabRef}}$\\
\\[-8pt]
$2s_{1/2}$ & 0.3745\hphantom{$^{\ref{FN:TabRef}}$} & 0.3533 & 0.3534 & 0.3534\hphantom{$^{\ref{FN:TabRef}}$}\\
& 0.3745$^{\ref{FN:TabRef}}$ & & & 0.3535$^{\ref{FN:TabRef}}$\\
\\[-8pt]
$2p_{1/2}$ & 0.0682\hphantom{$^{\ref{FN:TabRef}}$} & 0.0660 & 0.0660 & 0.0660\hphantom{$^{\ref{FN:TabRef}}$}\\
& 0.0683$^{\ref{FN:TabRef}}$ & & & 0.0661$^{\ref{FN:TabRef}}$\\
\\[-8pt]
$2p_{3/2}$ & 0.0094\hphantom{$^{\ref{FN:TabRef}}$} & 0.0094 & 0.0094 & 0.0093\hphantom{$^{\ref{FN:TabRef}}$}\\
& 0.0095$^{\ref{FN:TabRef}}$ & & & 0.0094$^{\ref{FN:TabRef}}$\\
\\[-6pt]
\colrule
\\[-6pt]
\multicolumn{5}{c}{\isotope[238][92]{U}}\\
\\[-6pt]
$1s_{1/2}$ & 5.4123\hphantom{$^{\ref{FN:TabRef}}$} & 4.9836 & 4.9860 & 4.9870\hphantom{$^{\ref{FN:TabRef}}$}\\
& 5.4124$^{\ref{FN:TabRef}}$& & & 4.9877$^{\ref{FN:TabRef}}$\\
\\[-8pt]
$2s_{1/2}$ & 0.8995\hphantom{$^{\ref{FN:TabRef}}$} & 0.8209 & 0.8213 & 0.8215\hphantom{$^{\ref{FN:TabRef}}$}\\
& 0.8995$^{\ref{FN:TabRef}}$ & & & 0.8216$^{\ref{FN:TabRef}}$ \\
\\[-8pt]
$2p_{1/2}$ & 0.2166\hphantom{$^{\ref{FN:TabRef}}$} & 0.2056 & 0.2056 & 0.2056\hphantom{$^{\ref{FN:TabRef}}$}\\
& 0.2166$^{\ref{FN:TabRef}}$ & & & 0.2057$^{\ref{FN:TabRef}}$ \\
\\[-8pt]
$2p_{3/2}$ & 0.0228\hphantom{$^{\ref{FN:TabRef}}$} & 0.0225 & 0.0225 & 0.0225\hphantom{$^{\ref{FN:TabRef}}$}\\
& 0.0228$^{\ref{FN:TabRef}}$ & & & 0.0226$^{\ref{FN:TabRef}}$ \\
\end{tabular}
\end{ruledtabular}
\caption{Nuclear model dependence of the all-order Wichmann and Kroll contribution to the Lamb shift in hydrogen-like \isotope[208][82]{Pb} ($R = \SI{5.5012}{\femto\metre}$) and \isotope[238][92]{U} ($R = \SI{5.8571}{\femto\metre}$) given in units of eV. Nuclear radii taken from Ref.~\cite{aam2013}.} \label{Tab:MDEl}
\end{table}
As seen from the table, there is a relatively large contribution from the finite nuclear size effects, in the low-lying states, which accounts for roughly \SI{0.1}{\eV} in lead and \SI{0.4}{\eV} in uranium for the 1s state. In contrast, the influence of the nuclear model is small and even the simple nuclear shell model differs by less than 0.1\% from the Fermi model. The results are in good agreement with previous calculations, see, \eg Ref.~\cite{yer2011, yas2015}.

\subsection{Muonic Atoms}
Having considered the VP in electronic atoms, we will now start our analysis of exotic atoms for the case of bound muons. Calculating the energy shift due to the electron-positron vacuum polarization can be again done with the help of Eq.~\eqref{Eq:EVP} by inserting the wave function of a bound muon instead of an electron for $\psi(\boldsymbol{r})$. In Table~\ref{Tab:MDMu}, we show the results for the Uehling and all-order Wichmann and Kroll contributions in muonic argon, xenon and uranium for the low-lying and the first few circular Rydberg states. Moreover, we indicate the dependence of the nuclear model in parentheses by displaying the absolute difference between the shown result obtained for a Fermi distribution and the result obtained from the homogeneous-sphere model. 
\begin{table*}
\begin{ruledtabular}
\begin{tabular}{cSSSSSSS}
& \multicolumn{2}{c}{$\mu^-$~\isotope[40][18]{Ar}} & \multicolumn{2}{c}{$\mu^-$~\isotope[132][54]{Xe}} & \multicolumn{2}{c}{$\mu^-$~\isotope[238][92]{U}}\\
State & \multicolumn{1}{c}{$E^{(1)}_\text{VP}$} & \multicolumn{1}{c}{$E^{(3+)}_\text{VP}$} & \multicolumn{1}{c}{$E^{(1)}_\text{VP}$} & \multicolumn{1}{c}{$E^{(3+)}_\text{VP}$} & \multicolumn{1}{c}{$E^{(1)}_\text{VP}$} & \multicolumn{1}{c}{$E^{(3+)}_\text{VP}$}\\
\\[-6pt]
\colrule
\\[-6pt]
$1s_{1/2}$ & -5533.3(97) & 3.2 & -39775.6(3308) & 137.3(9) & -74314.8(8166) & 698.3(47) \\
$2s_{1/2}$ & -821.9(12) & 0.7 & -8973.2(724) & 53.0(2) & -23266.2(2773) & 371.2(22) \\
$2p_{1/2}$ & -683.3(1) & 0.8 & -12616.6(161) & 74.3(1) & -39845.5(75) & 523.8(11) \\
$2p_{3/2}$ & -671.5(1) & 0.8 & -11701.5(208) & 71.5 & -36970.1(727) & 506.9(7)\\
$3d_{5/2}$ & -129.1 & 0.2 & -3217.9(2) & 31.5 & -13294.8(127) & 293.1(1) \\
$4f_{7/2}$ & -30.1 & 0.1 & -1104.2 & 14.9 & -5055.4(1) & 159.6(1) \\
$5g_{9/2}$ & -7.5 & 0.0 & -429.4 & 7.5 & -2203.1 & 90.2 \\
$6h_{11/2}$ & -1.9 & 0.0 & -178.3 & 3.9 & -1042.0 & 52.8 \\
$7i_{13/2}$ & -0.5 & 0.0 & -76.6 & 2.1 & -517.6 & 31.6 \\
$8k_{15/2}$ & -0.1 & 0.0 & -33.5 & 1.1 & -264.9 & 19.2 \\
$9l_{17/2}$ & 0.0 & 0.0 & -14.7 & 0.6 & -138.0 & 11.8 \\
$10m_{19/2}$ & 0.0 & 0.0 & -6.4 & 0.3 & -72.6 & 7.2 \\
$11n_{21/2}$ & 0.0 & 0.0 & -2.8 & 0.2 & -38.3 & 4.5 \\
$12o_{23/2}$ & 0.0 & 0.0 & -1.2 & 0.1 & -20.2 & 2.7 \\
$13q_{25/2}$ & 0.0 & 0.0 & -0.5 & 0.0 & -10.6 & 1.7 \\
$14r_{27/2}$ & 0.0 & 0.0 & -0.2 & 0.0 & -5.5 & 1.0 \\
\end{tabular}
\end{ruledtabular}
\caption{Uehling and all-order Wichmann and Kroll contribution to the Lamb shift in muonic \isotope[40][18]{Ar} ($R = \SI{3.4274}{\femto\metre}$), \isotope[132][54]{Xe} ($R = \SI{4.7859}{\femto\metre}$) and \isotope[238][92]{U} ($R = \SI{5.8571}{\femto\metre}$) given in units of eV. Calculations are performed for a Fermi model and the model dependence as the absolute difference to the homogenous sphere model is given in parentheses. Nuclear radii taken from Ref.~\cite{aam2013}.} \label{Tab:MDMu}
\end{table*}

As seen from the table, the effect of the vacuum polarization is strongly enhanced in muonic atoms and remains relatively large even for highly excited Rydberg states, due to the muon orbiting closer to the nucleus. Moreover, it can be seen that the low-lying states have a very large dependence on the nuclear model, especially for heavy elements, which limits the expected accuracy of these values. Additionally, the shifts of the low-lying states are also dependent on the nuclear charge radius, which has to be taken from experimental data and gives rise to additional uncertainty. However, this effect is much smaller than the dependence on the nuclear model. For example, for muonic uranium, varying the nuclear radius at the level of its experimental precision changes the Uehling and all-order Wichmann and Kroll shift of the $1s$ state by approximately \SI{42}{\eV} and \SI{0.4}{\eV}, respectively. In contrast to the low-lying states, the circular Rydberg states have almost no dependence on the structure of the nucleus. The observed nuclear model dependence of the higher-order VP correction is comparable in size to the influence of the nuclear model on the all-order self-energy, see Ref.~\cite{ore2022}. However, the dependence on the full all-order VP is naturally much larger due its larger overall size.

The presented results are in good agreement with the data available in the literature. For example, Rinker and Wilets~\cite{raw1975} find a value of $E^{(1)}_\text{VP} = \SI{-76225}{\eV}$ and $E^{(3+)}_\text{VP} = \SI{691}{\eV}$ for the $1s_{1/2}$ state in uranium and  $E^{(1)}_\text{VP} = \SI{-1043}{\eV}$ and $E^{(3+)}_\text{VP} = \SI{51}{\eV}$ for the $6h_{11/2}$ state. The values for the Uehling correction are systematically larger because they use a smaller radius of the uranium nucleus and the Wichman-Kroll correction is systematically smaller since the partial-wave expansion is terminated earlier in their computation.

\subsection{Antiprotonic Atoms}
In contrast to electrons and muons, antiprotons are composite particles that have a finite charge radius themselves. This will naturally modify the interaction potential between the nucleus and the antiproton. Assuming that both, the nucleus and the antiproton are homogeneously charged spheres with radii $R_1$ and $R_2$ ($R_1 > R_2$) whose centers are separated by $r$, it is easy to show~\cite{spbi2005} that the potential is given by:
\begin{widetext}
    \begin{equation} \label{Eq:FntPart}
        V(r) = \begin{cases}
-\frac{\alpha Z}{10R_1^3}(-5r^2+15R_1^2-3R_2^2) &\text{: $0\leq r \leq R_1-R_2$}\\
\frac{\alpha Z}{160rR_1^3R_2^3}[r^6-15r^4(R_1^2+R_2^2)+40r^3(R_1^3+R_2^3)-45r^2(R_1^2-R_2^2)^2 &\text{: $R_1-R_2 \leq r \leq R_1+R_2$}\\
\hphantom{\frac{\alpha Z}{160}}+24r(R_1+R_2)^3(R_1^2-3R_1R_2+R_2^2)-5(R_1-R_2)^4(R_1^2+4R_1R_2+R_2^2)] \\
-\frac{\alpha Z}{r} &\text{: $r \geq R_1+R_2$}
\end{cases}
\end{equation}
\end{widetext}
In order to calculate the wave function of the antiproton, we add the difference between the potential seen by a point charge in the presence of a homogeneously charged sphere and Eq.~\eqref{Eq:FntPart} into the Dirac equation. With these antiprotonic wave functions, we can calculate the vacuum polarization shift again using Eq.~\eqref{Eq:EVP}. The results for the same elements and atomic states as in the muonic case are shown in Table~\ref{Tab:MDpb} along with the nuclear model dependence.
\begin{table*}
\begin{ruledtabular}
\begin{tabular}{cSSSSSS}
& \multicolumn{2}{c}{$\bar{p}$~\isotope[40][18]{Ar}} & \multicolumn{2}{c}{$\bar{p}$~\isotope[132][54]{Xe}} & \multicolumn{2}{c}{$\bar{p}$~\isotope[238][92]{U}}\\
State & \multicolumn{1}{c}{$E^{(1)}_\text{VP}$} & \multicolumn{1}{c}{$E^{(3+)}_\text{VP}$} & \multicolumn{1}{c}{$E^{(1)}_\text{VP}$} & \multicolumn{1}{c}{$E^{(3+)}_\text{VP}$} & \multicolumn{1}{c}{$E^{(1)}_\text{VP}$} & \multicolumn{1}{c}{$E^{(3+)}_\text{VP}$}\\
\\[-6pt]
\colrule
\\[-6pt]
$1s_{1/2}$ & -29835.9(9174) & 8.5(2) & -80418.4(26424) & 192.3(29) & -112104.3(29878) & 841.4(88) \\
$2s_{1/2}$ & -9349.4(2759) & 4.5(1) & -48711.4(7543) & 154.8(12) & -85685.0(8933) & 751.1(46) \\
$2p_{1/2}$ & -14715.5(715) & 6.1 & -65659.0(10361) & 176.3(20) & -100270.9(19323) & 799.0(78) \\
$2p_{3/2}$ & -14671.2(740) & 6.1 & -65572.3(10228) & 176.2(20) & -100201.0(19229) & 798.7(78) \\
$3d_{5/2}$ & -5698.4(136) & 3.6 & -49082.1(1217) & 156.4(7) & -87082.1(8105) & 753.7(57) \\
$4f_{7/2}$ & -2383.4(1) & 2.1 & -32277.0(3114) & 130.3 & -72352.6(1269) & 701.4(28) \\
$5g_{9/2}$ & -1140.3 & 1.2 & -18912.3(638) & 100.0 & -56049.3(5300) & 634.6(2) \\
$6h_{11/2}$ & -594.7 & 0.8 & -11058.8(29) & 73.8 & -39655.3(3108) & 548.1(4)\\
$7i_{13/2}$ & -328.0 & 0.5 & -6820.9 & 54.5 & -26381.0(553)  & 451.2(2) \\
$8k_{15/2}$ & -187.8 & 0.3 & -4405.0 & 40.8 & -17643.0(32) & 363.4(3) \\
$9l_{17/2}$ & -110.3 & 0.2 & -2943.7 & 30.1 & -12174.8(1) & 292.1(2) \\
$10m_{19/2}$ & -65.9 & 0.2 & -2018.8 & 23.8 & -8641.9(1) & 235.7(1) \\
$11n_{21/2}$ & -39.9 & 0.1 & -1412.6 & 18.5 & -6272.6 & 191.3(1) \\
$12o_{23/2}$ & -24.3 & 0.1 & -1004.2 & 14.4 & -4634.8 & 156.1(1) \\
$13q_{25/2}$ & -14.9 & 0.1 & -722.9 & 11.4 & -3474.6 & 128.0 \\
$14r_{27/2}$ & -9.1 & 0.0 & -525.7 & 9.0 & -2636.1 & 105.6 \\
\end{tabular}
\end{ruledtabular}
\caption{Uehling and all-order Wichmann an Kroll contribution to the Lamb shift in antiprotonic \isotope[40][18]{Ar} ($R = \SI{3.4274}{\femto\metre}$), \isotope[132][54]{Xe} ($R = \SI{4.7859}{\femto\metre}$) and \isotope[238][92]{U} ($R = \SI{5.8571}{\femto\metre}$) given in units of eV. Calculations are performed for a Fermi model and the model dependence as the absolute difference to the homogenous sphere model is given in parentheses. Nuclear radii taken from Ref.~\cite{aam2013}.} \label{Tab:MDpb}
\end{table*}

The numerical results in the table clearly show that the effect of VP is even more enhanced in antiprotonic atoms compared to that of muonic ones because of their  higher mass. In the high-$Z$ regime, the VP shift is very similar in size for the first few low-lying states due to the wave function being almost fully contained in the nucleus where the VP potential is close to being constant. Moreover, the correction remains very large for the circular Rydberg states at more than \SI{2}{\kilo\eV} for the Uehling part and more than \SI{100}{\eV} for the all-order Wichmann and Kroll contribution in antiprotonic uranium at the $14r_{27/2}$ state. Despite this fact, the influence of the nuclear model is relatively small in the highly excited states while its influence is naturally very large for the lowest states. As for muonic atoms, the uncertainty coming from the nuclear charge radius is smaller than the nuclear model dependence at around \SI{81}{\eV} and \SI{0.5}{\eV} for $E^{(1)}_\text{VP}$ and $E^{(3+)}_\text{VP}$ in the $1s$ state of uranium, reducing quickly for higher states. The influence of the finite size of the antiproton on the VP is relatively small and amounts to only around \SI{30}{\eV} and \SI{1}{\eV} of the Uehling and Wichmann and Kroll correction for the low-lying states of uranium and is even smaller for lighter elements and highly excited states. The results are in good agreement with Patkó$\Check{\text{s}}$ and Pachucki~\cite{pap2025}, who find a shift of 909.72~\si{\eV} for the $10m_{19/2} \to 9l_{17/2}$ transition in \isotope[132][54]{Xe} by including the one- and two-loop VP in the Schrödinger equation, compared to our value for the one-loop VP of 918.6~\si{\eV}.

\section{Conclusion} \label{Sec:Conc}
In conclusion, we have presented a theoretical study of the VP contribution to the Lamb shift. We have considered the lowest-order (in $\alpha Z$) Uehling and all-order Wichmann and Kroll corrections in electronic, muonic and antiprotonic atoms. Moreover, we have explicitly investigated the influence of the nuclear model and have performed calculations for a wide range of elements and atomic states, including circular Rydberg states. Additionally, we examined the influence of the finite size of the antiproton on the VP. Our calculations have shown that the lowest- and higher-order VP corrections are strongly enhanced in exotic atoms and remain relatively large for highly excited Rydberg states. The influence of the nuclear structure is much larger in exotic atoms compared to electronic ones, but becomes negligible in highly excited states. The influence of the finite size of the antiproton on the VP is only relevant for the lowest states and only accounts for less than 0.1\% of the VP shift even in the strongest cases.

In this work, we only considered the VP correction for a single bound electron, muon or antiproton without screening from residual electrons. In experiments, the screening corrections to the VP are expected to be small for most transitions in exotic ions since almost all residual electrons are ejected during the cascade. Moreover, we have not considered the shift due to two-loop VP effects. Including the all-order VP potential into the Dirac equation shows that the higher-order VP corrections can be on the order of a few \si{\eV} for exotic Rydberg states.

\begin{acknowledgments}
This work was supported by the German Research Foundation (Deutsche Forschungsgemeinschaft, DFG) under Project 546193616. Paul Indelicato is a member of the Helmholtz Alliance HA216/EMMI. P.I thanks the Institute of Physics of CNRS for financing the high-performance computer used to perform this work.
\end{acknowledgments}

\appendix

\section{Numerical Solution of the Dirac Equation on a Grid} \label{Ap:Dirac}
To solve the Dirac equation on a grid, we introduce the variables $x = (r-r_a)/h$ and $h = r_b - r_a$ for each interval $[r_a, r_b]$ with given boundary conditions at $r_a$, and write the Dirac equation as
\begin{equation} \label{Eq:DiracInt}
\begin{aligned}
     (xh+r_a)G'_x+\kappa hG + UhF - (xh+r_a)hF &= 0~,\\
     (xh+r_a)F'_x-\kappa hF - UhG - (xh+r_a)hG &= 0~,
\end{aligned}
\end{equation}
where $U = r(V(r)-E)$ and $G = rF^1_\kappa$ and $F = rF^2_\kappa$ are the upper and lower components of the solutions of the radial Dirac equation. The size of the intervals is chosen small enough so that the potential can be represented by a cubic polynomial $U = \sum_{k=0}^3 u_kx^k$ with sufficient precision. We can then find the solutions as a power series in x
\begin{equation} \label{Eq:GFPS}
    \begin{aligned}
        G(x) &= \sum_{n=0}^{n_\text{max}} a_n x^n~, \\
        F(x) &= \sum_{n=0}^{n_\text{max}} b_n x^n~,
    \end{aligned}
\end{equation}
with $a_0$ and $b_0$ being determined by the boundary condition at $r_a$ and the sum of all coefficients determining the boundary condition for the next interval at $r_b$. If the desired numerical precision is not reached by summing up to $n_\text{max}$, we divide the interval in the middle into two sub-intervals. A recursion relation for the coefficients can be found by inserting the power expansions~\eqref{Eq:AOVP} into Eq.~\eqref{Eq:DiracInt}. For $r_a \neq 0$, one obtains
\begin{equation}
    \begin{aligned}
        a_n =& -\frac{h}{nr_a}[(n-1+\kappa)a_{n-1}+(u_0-r_a)b_{n-1}\\
        &+(u_1-h)b_{n-2}+u_2b_{n-3}+u_3b_{n-4}]~,\\
        b_n =& \frac{h}{nr_a}[(-n+1+\kappa)b_{n-1}+(u_0+r_a)a_{n-1}\\
        &+(u_1+h)a_{n-2}+u_2a_{n-3}+u_3a_{n-4}]~.
    \end{aligned}
\end{equation}
For the case $r_a = 0$, the solutions are represented by
\begin{equation} \label{Eq:GFPSra0}
    \begin{aligned}
        G(x) &= x^s\sum_{n=0}^{n_\text{max}} a_n x^n~, \\
        F(x) &= x^{s+t}\sum_{n=0}^{n_\text{max}} b_n x^n~,
    \end{aligned}
\end{equation}
and we need to distinguish two cases for regular potentials with $u_0 = 0$. For the case $u_0 = 0$, $\kappa < 0$, we take $s = \vert \kappa \vert$, $t = 1$ and
\begin{equation}
    a_0 = 1~, b_0 = \frac{h+u_1}{1+2\vert \kappa \vert}~.
\end{equation}
The recursion relations take the form
\begin{equation}
\begin{aligned}
    -na_n&=(u_1-h)b_{n-2}+u_2b_{n-3}+u_3b_{n-4}~,  \\
    (2\vert \kappa \vert+n+1)b_n&=(u_1+h)a_{n}+u_2a_{n-1}+u_3a_{n-2}~.  \\
\end{aligned}
\end{equation}
Finally, for $u_0 = 0$, $\kappa > 0$ we have $s = \kappa + 1$, $t=-1$,
\begin{equation}
    a_0 = \frac{h-u_1}{1+2\kappa}~, b_0 = 1~,
\end{equation}
and
\begin{equation}
\begin{aligned}
    (2\kappa+n+1)a_n&=(h-u_1)b_{n}-u_2b_{n-1}-u_3b_{n-2}~,  \\
    nb_n&=(h+u_1)a_{n-2}+u_2a_{n-3}+u_3a_{n-4}~,  \\
\end{aligned}
\end{equation}
see Ref.~\cite{yer2011} for further details.

\section{Analytical Expression for the One-Interaction Term}  \label{Ap:I1Ana}
For the sphere and shell model, we can split the integral into three regions
\begin{equation}
\begin{aligned}
    \int_0^\infty &dr'~r'^2V(r') \sum_{n,m=1}^2 [F^{nm}_\kappa(r,r',iu)]^2 \equiv \int_0^\infty dr'~A(r') \\
    &= \int_0^{a_1} dr'~A(r')+\int_{a_1}^{a_2} dr'~A(r')+\int_{a_2}^{\infty} dr'~A(r') \\
    &\equiv T_1+T_2+T_3~,
\end{aligned}
\end{equation}
where $a_1 = \text{min}(r,R)$ and $a_2 = \text{max}(r,R)$. If the potential has the form $V(r) = V_0 + V_2r^2$ for $r \leq R$ and $V(r) = V_\infty/r$ for $r > R$, we can use the analytical form of the free electron Green's function to define
\begin{equation}
\begin{aligned}
    T_1 &= V_0 T_1^2 +V_2T_1^4~, \\
    T_2 &=
    \begin{cases}
        V_0 T_2^2 +V_2T_2^4 &\text{for}~r\leq R~, \\
        V_\infty \widetilde{T}_2^1 &\text{for}~r>R~,\\
    \end{cases}\\
    T_3 &= V_\infty T_3^1~.
\end{aligned}
\end{equation}
Using the functions
\begin{equation} \label{Eq:BInt}
    \begin{aligned}
        J_{k,n}(a) &= \int_0^a dr'~r'^k (j_n(icr'))^2~,\\
        H_{k,n}(a) &= \int_a^\infty dr'~r'^k (h^{(1)}_n(icr'))^2~,\\
    \end{aligned}
\end{equation}
where $j_n(x)$ is the spherical Bessel function and $h^{(1)}_n(x)$ is the spherical Hankel function of the first kind, and $l = \vert \kappa\vert$, $m = l-1$, we can find the solutions
\begin{widetext}
\begin{equation}
    \begin{aligned}
        T_1^k =& 2(1-u^2)c^2 [(h_l^{(1)}(icr))^2J_{k,l}(a_1)+(h_m^{(1)}(icr))^2J_{k,m}(a_1)]-2c^4 [(h_m^{(1)}(icr))^2J_{k,l}(a_1)+(h_l^{(1)}(icr))^2J_{k,m}(a_1)]~,\\
        T_2^k =&2(1-u^2)c^2 [(j_l(icr))^2(H_{k,l}(a_1)-H_{k,l}(a_2))+(j_m(icr))^2(H_{k,m}(a_1)-H_{k,m}(a_2))]\\&-2c^4 [(j_m(icr))^2(H_{k,l}(a_1)-H_{k,l}(a_2))+(j_l(icr))^2(H_{k,m}(a_1)-H_{k,m}(a_2))]~,\\
        \widetilde{T}_2^k =&2(1-u^2)c^2 [(h_l^{(1)}(icr))^2(J_{k,l}(a_2)-J_{k,l}(a_1))+(h_m^{(1)}(icr))^2(J_{k,m}(a_2)-J_{k,m}(a_1))]\\&-2c^4 [(h_m^{(1)}(icr))^2(J_{k,l}(a_2)-J_{k,l}(a_1))+(h_l^{(1)}(icr))^2(J_{k,m}(a_2)-J_{k,m}(a_1))]~,\\
        T_3^k =& 2(1-u^2)c^2 [(j_l(icr))^2H_{k,l}(a_2)+(j_m(icr))^2H_{k,m}(a_2)]-2c^4 [(j_m(icr))^2H_{k,l}(a_2)+(j_l(icr))^2H_{k,m}(a_2)]~.
    \end{aligned}
\end{equation}    
\end{widetext}
If the potential has a more complicated form in the inner region, like for the Fermi model, we integrate the expressions for $T_1$ and $T_2$ numerically instead until $R_0$, but we can still use the analytical form of $T_3$ and $\widetilde{T}_2^1$.

Analytical expressions for the integrals in Eq.~\eqref{Eq:BInt} can be found in terms of hypergeometric functions for $J_{k,n}$ and in terms of incomplete Gamma functions for $H_{k,n}$ using the finite asymptotic expansion of $h^{(1)}_n$. We use the arb C library~\cite{joh2017} to calculate these special functions with sufficient precision.


\bibliography{refs2022,refs2023,refs2024,refs2025}

\end{document}